\documentclass[fleqn,12pt,twoside,pdflatex]{article}
\usepackage{espcrc1}
\usepackage[english]{babel}
\usepackage{amssymb}

\usepackage{graphicx}
\usepackage[figuresright]{rotating}

\newcommand{\AmS}{{\protect\the\textfont2
  A\kern-.1667em\lower.5ex\hbox{M}\kern-.125emS}}

\title{Segregation phases in a vibrated binary granular layer}

\author{P.M. Reis\address{Manchester Center for Nonlinear Dynamics,
        Department of Physics and Astronomy, \\University of
        Manchester,
        Oxford Road, Manchester M13 9PL, UK}\thanks{email address: pedro@reynolds.ph.man.ac.uk},
        T.
        Mullin\addressmark \
        and G. Ehrhardt\address{Theoretical Physics Group,
        Department of Physics and Astronomy,\\ University of Manchester, Oxford Road, Manchester M13 9PL, UK}
        }

\begin{document}

\maketitle

\begin{abstract}
We present the results of an experimental study of patterned
segregation in a horizontally shaken shallow layer of a binary
mixture of dry particles. As the compacity, $C$, of the mixture
was increased, the evolution of three distinct phases was
observed. We classify them as binary gas, segregation liquid and
segregation crystal phases using macroscopic and microscopic
measures. The binary gas to segregation liquid transition is
consistent with a continuous phase transition and includes the
characteristic feature of critical slowing down. At high
compacities we observed an intriguing slow oscillatory state.

\end{abstract}

\section{Introduction}
\label{sec:introduction}

In segregation, excitation via flow or shaking can,
counter-intuitively, cause an initially homogeneous mixture of
grains to de-mix \cite{mullin:2002,shinbrot:2000}. Intriguingly,
it does not always happen and the conditions for its occurrence
are difficult to predict. Hence, a better understanding would have
a major economic impact in the pharmaceutical, chemical processing
and civil engineering industries \cite{williams:1976}. However,
despite decades of research, a predictive model of the process has
yet to emerge. More recently, segregation has received
considerable attention from the physics community as an example of
a challenging far from equilibrium system with pattern formation
\cite{shinbrot:2000}. Although many small scale laboratory studies
involving vibration \cite{burtally:2002}, avalanching in partially
filled horizontal rotating drums \cite{gray:2001} and
stratification in vertically poured mixtures \cite{makse:1997}
have been investigated and a variety of mechanisms proposed
\cite{ottino:2000}, an understanding of the fundamental principles
involved remains incomplete.

In our experiments
\cite{mullin:2000,mullin:2002,reis:2002,reis:2003b} we have
focused on quasi-2-dimensional layers of binary mixtures of
particles which are horizontally driven by the stick--slip
frictional interaction with the surface of a horizontal tray.
Vibration of an initially homogeneous mixed binary layer can give
rise to robust segregation of the constituents, depending on the
total filling fraction of the layer. We call this parameter the
\emph{compacity}. As the compacity was increased the evolution of
distinct phases was observed. We identified them as a \emph{binary
gas}, a \emph{segregation liquid} and a \emph{segregation crystal}
using both microscopic and macroscopic measures to identify their
properties. This horizontal and quasi-2-dimensional set up gives
the practical advantage that any collective behavior is readily
visualized, vertical compaction effects are effectively eliminated
and the material is in contact with the forcing throughout the
drive cycle.

For this class of quasi-2D binary granular systems a qualitative
segregation mechanism has been suggested
\cite{reis:2003b,aumaitre:2001,shinbrot:2001,duran:1999} using the
idea of \emph{excluded volume depletion} as in colloidal systems
and binary alloys \cite{hill:1994}. This is in the spirit of the
physics of complex systems where attempts have been made to unify
descriptions of granular materials, colloids, gels and foams
\cite{liu:1998,trappe:2001}. We discuss the relevance of these
ideas to our observations in some concluding remarks.

\section{The experiment}
\label{sec:experiment}

\begin{figure}[h]
        \begin{center}
            \includegraphics[width=14cm]{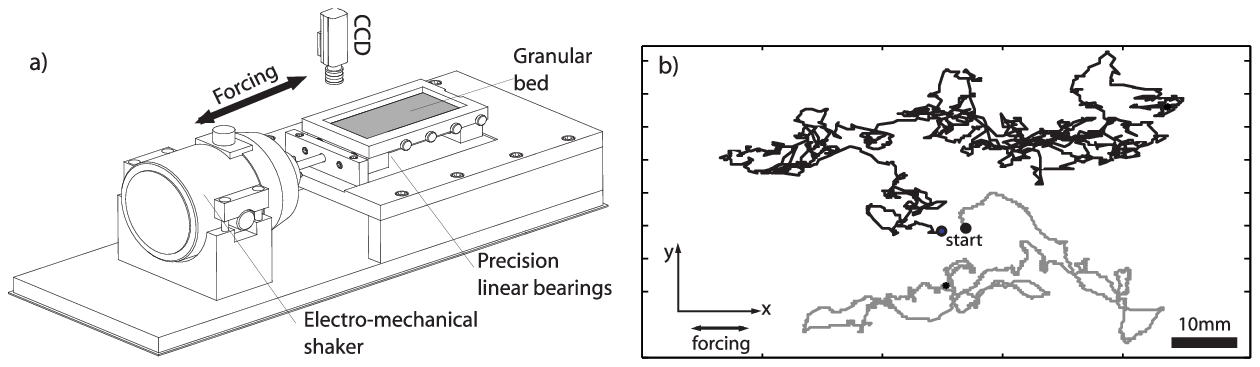}\\
            \includegraphics[width=14cm]{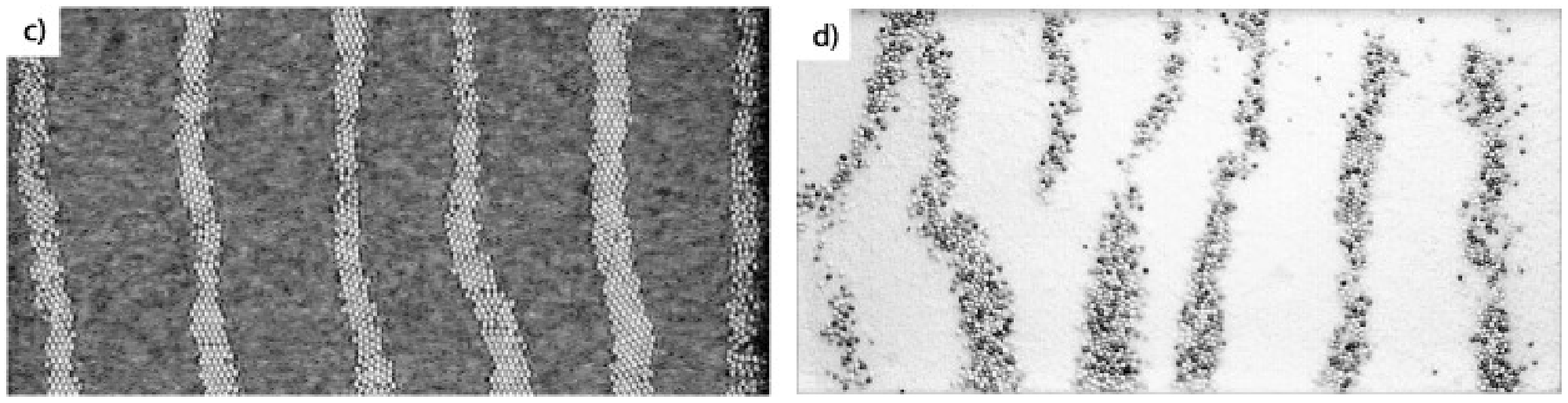}
        \end{center}

        \caption{\footnotesize
        (a) Schematic diagram of  the experimental
        apparatus.
        (b) Trajectories of two individual particles over $1min$: black line corresponds to poppy
        seed and grey line to phosphor-bronze sphere.
        (c) Typical segregation pattern for mixture A: poppy seeds
        (grey regions)
        and phosphor-bronze spheres (white regions).
         (d) Typical segregation pattern for mixture B: sugar particles (shaded grey regions) and polystyrene
         spheres (white regions). Both patterns self-organised $15min$ after starting from a homogeneous mixture.}
        \label{fig:apparatus_patterns}
    \end{figure}

A schematic diagram of the top view of the apparatus is presented
in Fig. \ref{fig:apparatus_patterns}(a). It consisted of an
accurately levelled and horizontal smooth rectangular tray, of
dimensions $(x,y)=180\times90mm$ with a flatness of less than
$\pm5\mu m$, on which particles were vibrated longitudinally. The
tray was made from aluminum tool plate for increased rigidity and
was mounted on a horizontal platform which was connected to an
electro-mechanical shaker. Its motion was a close approximation to
one-dimensional as it was constrained by four lateral high
precision linear bush bearings. The dynamic displacement and
acceleration of the shaking bed was monitored by a Linear
Displacement Variable Transformer (LVDT) and a piezoelectric
accelerometer. The tray's acceleration was checked to be
sinusoidal, to 0.1\% in power, using the power spectrum of the
accelerometer trace.

We have investigated segregation using a variety of mixtures but
here focus on results for two specific types: \emph{mixture A}
(polystyrene spheres + sugar particles)  and \emph{mixture B}
(phosphor-bronze spheres + poppy seeds). In particular, we have
used mixture B for the detailed quantitative results. The details
for the particles' material properties is presented in Table 1.

At this point we define the \emph{layer compacity} to be the total
filling fraction of the system,
\begin{equation}
C=\frac{N_{1}A_{1}+N_{2}A_{2}}{xy}, \label{eqn:compacity}
\end{equation}
where $N_{1}$ and $N_{2}$ are the numbers of species $a$ and $b$
in the layer, $A_1$ and $A_2$ are the two dimensional projected
areas of the respective individual particles
($A_1=(0.87\pm0.15)mm^2$ for poppy seeds and
$A_2=(1.77\pm0.06)mm^2$ for phosphor-bronze spheres) and $x$ and
$y$ are the longitudinal and transverse dimensions of the
rectangular tray. We have chosen to fix the forcing at a single
frequency $f=12Hz$ and amplitude $A=\pm1.74mm$ and explore the
effect of the layer compacity.

    \begin{table}[t]
        \begin{center}
        \begin{tabular}{c|c|c|c}

Material    & Avg. Diameter ($mm$)   & Density ($gcm^{-3}$) &
Shape
\\ \hline

poppy seed &1.07 ($\pm17\%$)   & 0.2  & flat 'kidney' shaped
\\ \hline

phosphor-bronze&1.50 & 8.8  & precision spherical \\ \hline

polystyrene    &0.5  & 1.1 & precision spherical \\ \hline

sugar &1.71& 1.6  & roughly spherical

        \end{tabular}
        \label{tab:properties}
    \caption{Physical properties of the particles used.}
    \end{center}
    \end{table}

Throughout this study, experiments were performed in a
quasi-monolayer regime so that the smaller particles were rarely
on top of the larger ones. This was particularly the case for the
mixtures of poppy seeds and phosphor bronze spheres since the size
ratio for these two types of particle is $\sim2:3$. However, some
overlapping of particles did occur so that compacities up to
$C\sim 1.1$ were achieved in the high packing limit.

\section{Single particle motion}
\label{sec:singleparticle}

As discussed above, the forcing was strictly sinusoidal but both
types of particles were driven by a stick-and-slip interaction
with the oscillatory surface of the container. These interactions
induced stochastic motion and trajectories of single particles
with the characteristics of a 2D random walk were obtained.
Typical one minute long trajectories for a poppy seed and a
phosphor-bronze sphere are shown in Fig.
\ref{fig:apparatus_patterns}(b). The statistics of the motion were
anisotropic with a preference for motion in the direction of the
forcing. Note that the two types of particles responded to the
driving differently, due to their mass, shape and surface
properties. When many particles were present, collective motion
dominated with diffusion biased in the direction of the forcing.

\section{Granular segregation}
\label{sec:segregation}

  \begin{figure}[t]
        \begin{center}
           \includegraphics[height=6.5cm]{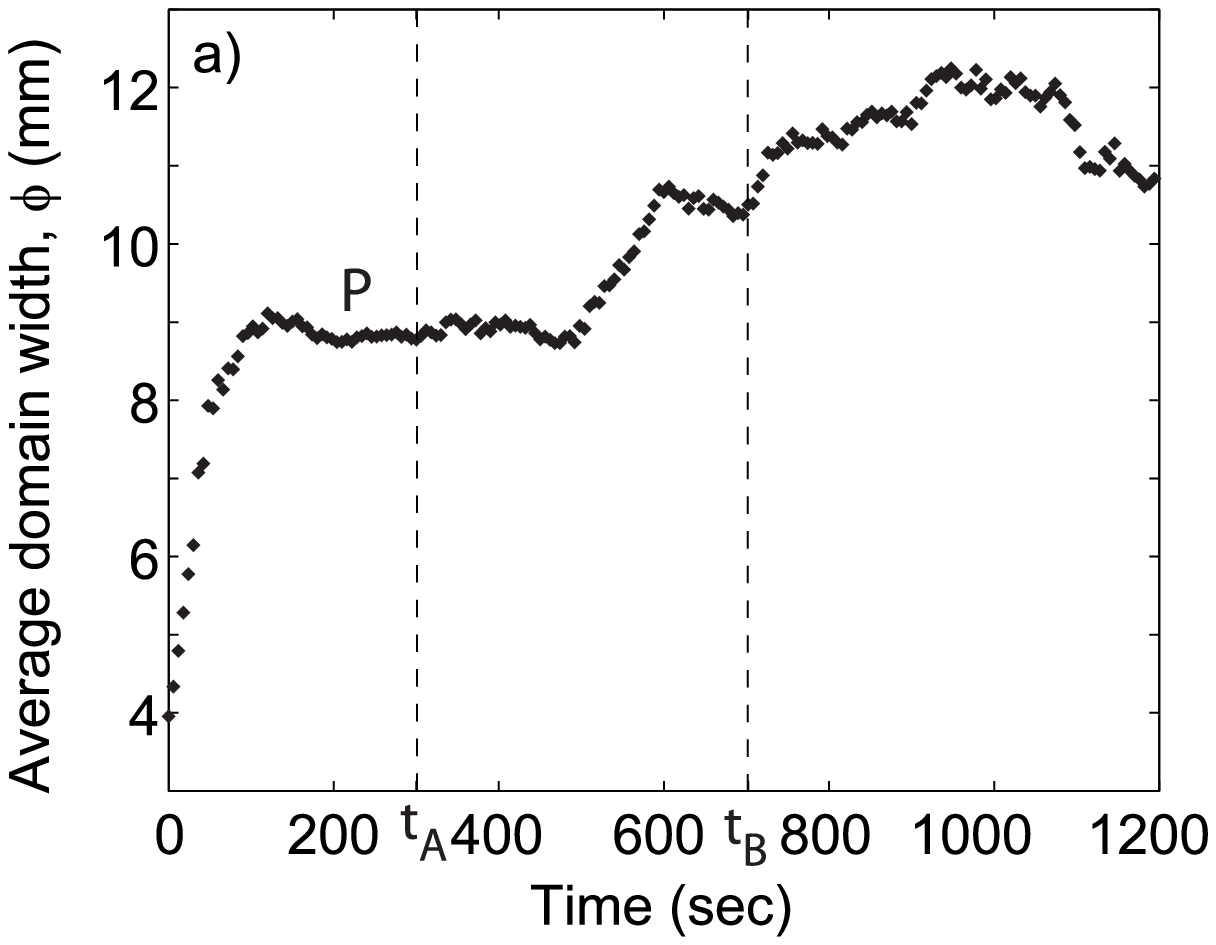}\includegraphics[height=6.5cm]{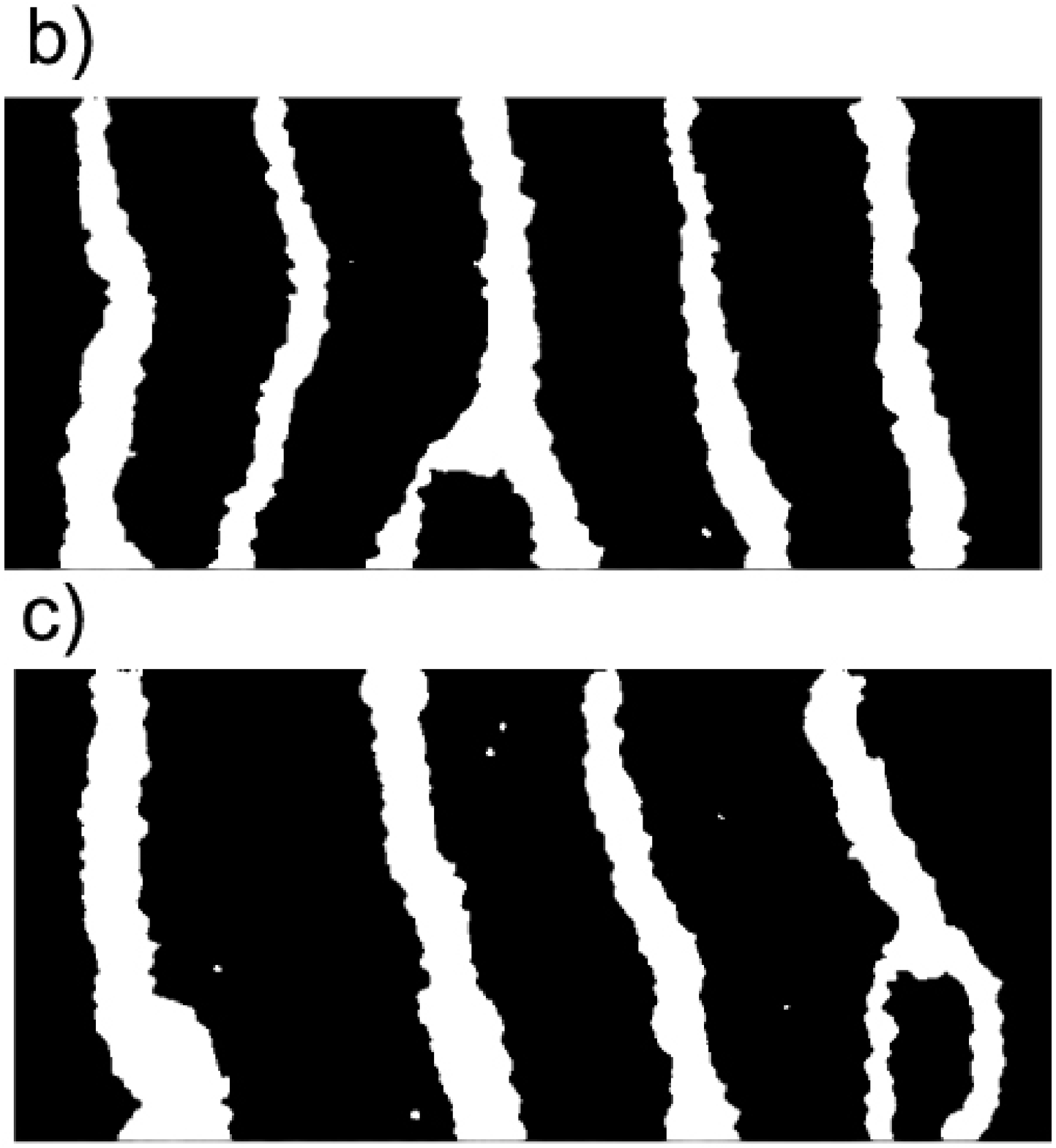}
        \end{center}
        \caption{ \footnotesize
        (a) Time series of the average domain width, $\phi$, over
        20min, for $C=1.057$. (b)\&(c) Binary frames of the segregation
        patterns (regions of poppy seeds in black and regions
        of phosphor-bronze spheres in white) at (b) $t_A=300sec$ and (c)
        $t_B=702sec$.}
        \label{fig:timeseries_file29}
    \end{figure}

In Fig. \ref{fig:apparatus_patterns}(c) and (d) we present two
examples of segregation patterns for the two mixtures considered.
The snapshots correspond to the segregated states which formed
after the mixture was vibrated for a period of 15 minutes. The
experimental runs were systematically started from a homogeneous
mixed layer using the following method. Firstly, $N_{2}$ small
particles were vibrated at large amplitudes, $A\sim\pm 5mm$,
creating an homogeneous and isotropic sub-monolayer. The large
particles were then suspended above the layer, on a horizontal
perforated plate with ($57\times28$) $2mm$ diameter holes arranged
in a triangular lattice. A shutter was then opened and the 1596
particles fell onto the layer of small particles, creating a near
homogeneous binary mixture. Domains such as those shown in Fig.
\ref{fig:apparatus_patterns}(c) and (d) then formed. The
differences in the physical properties of the two mixtures
illustrates that, in our system, patterned segregation is a robust
phenomena.

    \begin{figure}[t]
        \begin{center}
           \includegraphics[width=7.75cm]{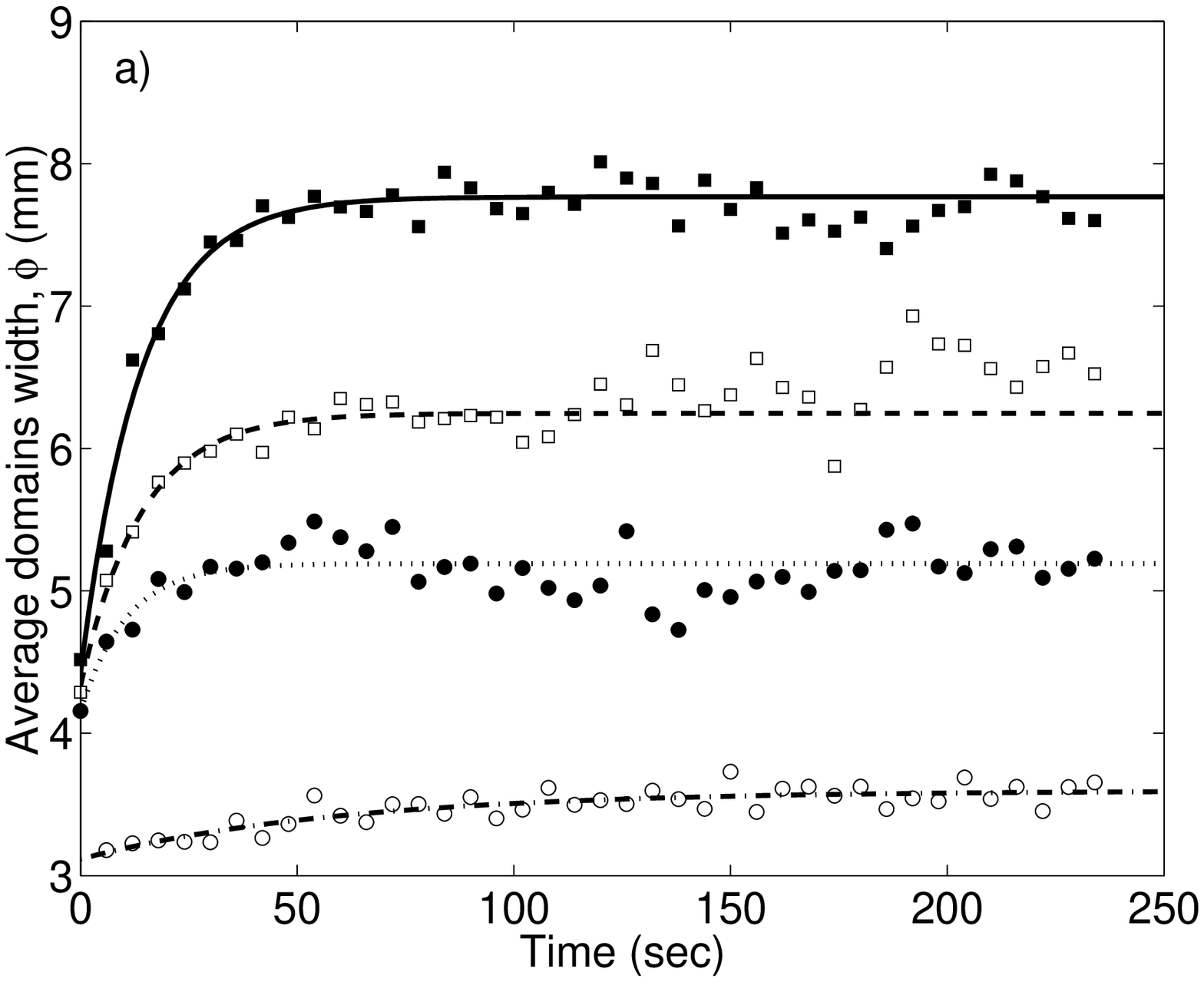}
            \includegraphics[width=7.75cm]{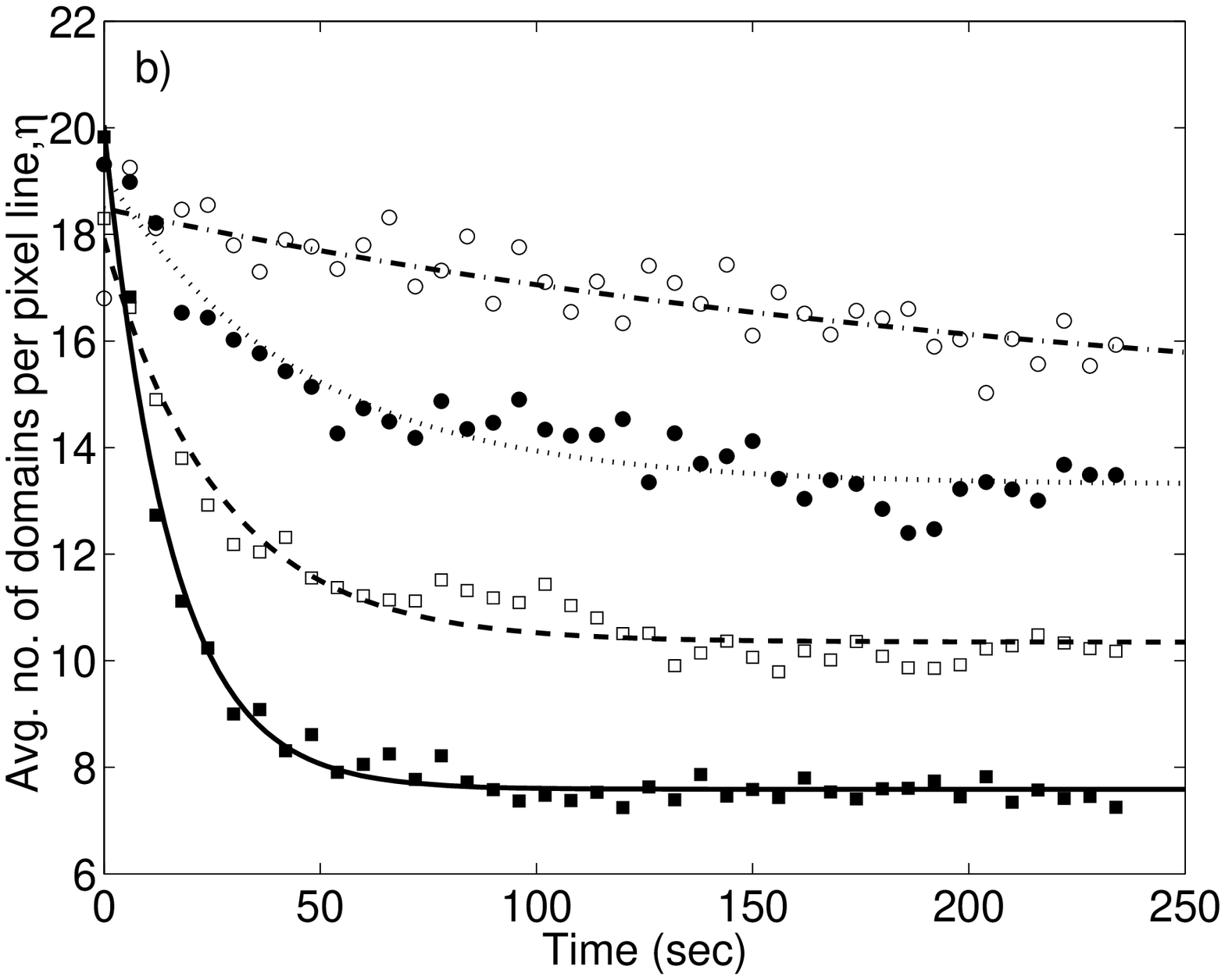}
        \end{center}
        \caption{ \footnotesize
        (a) Time-series of the average domain width, $\phi$.
        (b) Time-series for number of domains, $\eta$.
        ({\tiny{$\square$}})
         $C=0.873$,
        ({\tiny{$\blacksquare$}})  $C=0.729$ ,
        ($\bullet$)  $C=0.708$ , ($\circ$)  $C=0.602$. Lines are fits
        to Eqn. (\ref{eqn:fits}). All runs were started from a
        homogeneously mixed layer.}
        \label{fig:timeseries}
    \end{figure}

In order to quantify the dynamics of growth of the segregated
domains we focus on mixture B and establish two macroscopic
measures: the average longitudinal width and the average number of
domains of phosphor-bronze spheres. At this stage, the images of
the whole layer ($300\times600$ pixels) were processed such that
regions of poppy seeds were given a value of 0 (black) and regions
of spheres a value of 1 (white). For each acquired frame this was
calculated by scanning through each of the 300 horizontal lines
and the length, $L_i$, of each of the \emph{white steps} was
measured. We therefore define, $\phi=\langle L \rangle$, to be the
average value of the domain width and the average number of
domains, $\eta$, to be the number of such steps, per pixel line.

A typical time evolution of $\phi$ for $C=1.057$, over a period of
20min, is presented in Fig. \ref{fig:timeseries_file29}(a). It is
characterised by a fast initial segregation growth, over
timescales of the order of $\sim~1min$. During this period the
mechanism of the segregation process is as follows. Initially,
single large particles diffuse in a sea of the smaller ones. When
two large particles come close together, the smaller particles
cannot fit between them, and hence the pair is subjected to an
asymmetric pressure that will keep it together. Subsequently,
pairs may encounter others so that progressively larger clusters
form. The unidirectionality of the driving induces an asymmetry in
the segregated domains such that elongated domains of the larger
particles (phosphor-bronze spheres or sugar particles) develop.
The segregation domains were elongated in a direction
perpendicular to the drive. The growth of the domains then
exhibits an intermediate saturation level as can be seen in the
plateau labelled $P$ in Fig. \ref{fig:timeseries_file29}(a).

Further coarsening of the domains can take place, albeit at longer
timescales. Two examples of the patterns at later times are
presented in Fig. \ref{fig:timeseries_file29}(b) and (c) at
$t_A=300s$ and $t_B=702s$, respectively. During this period
coarsening has occurred via the break-up of one of the stripes and
consequent merging with its neighbours. In contrast to the fast
initial segregation, this slow coarsening may be associated with
large scale collective motion in the layer of poppy seeds.

Since there is a distinct separation of the timescales we will now
focus on the \emph{fast} initial segregation growth regime. In
Fig. \ref{fig:timeseries}(a) and (b), we present typical
time-series for $\phi$ and $\eta$, respectively, for four
different values of $C$, over the first $4min$ of vibration. The
superimposed lines are fits of,
        \begin{equation}
            \phi (C,t)=\Sigma^{\phi} -
            b^{\phi} .\exp\left(-\frac{t}{t^{\phi} _s} \right)
\quad \mathrm{and} \quad \eta (C,t)=\Sigma^{\eta} +
            b^{\eta} .\exp\left(-\frac{t}{t^{\eta} _s} \right),
            \label{eqn:fits}
        \end{equation}
to the experimental time series of both $\phi$ and $\eta$, where
$\Sigma^{\phi}$ and $\Sigma^{\eta}$ are the respective saturation
levels and $t^{\phi}$ and $t^{\eta} _s$, the \emph{segregation
times}, are timescales associated with the saturation of the
domain growth. We denote $\Sigma^\phi$ by the \emph{segregation
level}. A result which may be inferred from both sets of time
sequences in Fig. \ref{fig:timeseries} is that, when the domains
form, there is a progressive coalescence of clusters with an
increase of the average width and corresponding decrease in the
number of domains in the system.

\section{Segregation Phase transition}
\label{sec:transition}

We next investigated the dependence of the saturation levels,
$\Sigma$, and segregation times, $t_s$, on the compacity of the
layer for both $\phi$ and $\eta$. This was done by incrementally
increasing the number of poppy seeds, $N_{1}$, in the layer, in
measured steps, while the number of phosphor-bronze spheres,
$N_{2}=1596$, held constant, i.e. $C=C(N_{1})$. We chose to
perform the experiments by changing the numbers of poppy seeds.
This can be regarded analogous to the experimental protocol in
colloid-polymer mixtures of increasing the concentration of the
polymer in solution (see Section 8 and Ref. \cite{poon:2002}). The
qualitative behaviour presented here was found to be robust over a
range of $N_2$.

    \begin{figure}[t]
        \begin{center}
            \includegraphics[width=7.75cm]{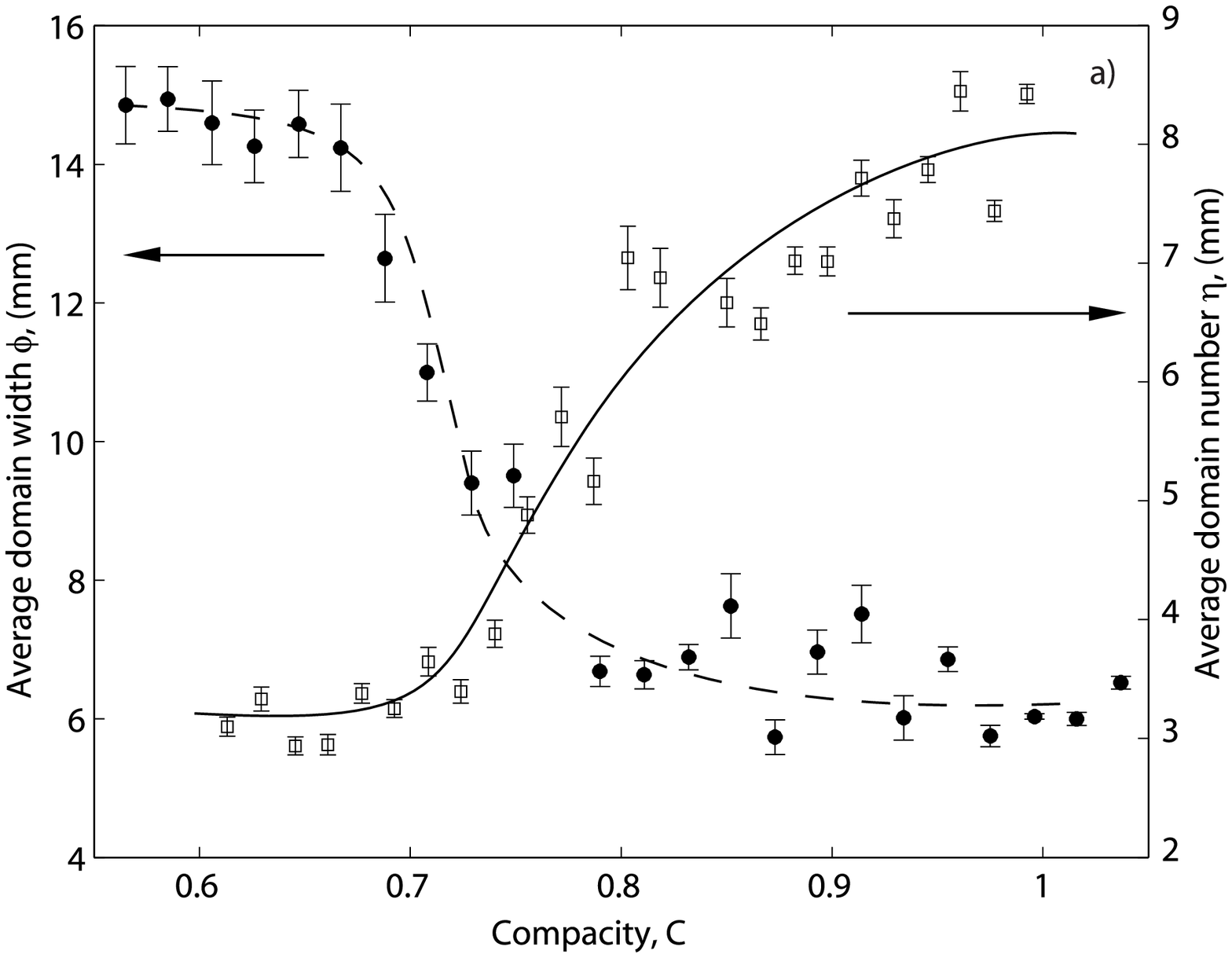}
            \includegraphics[width=7.75cm]{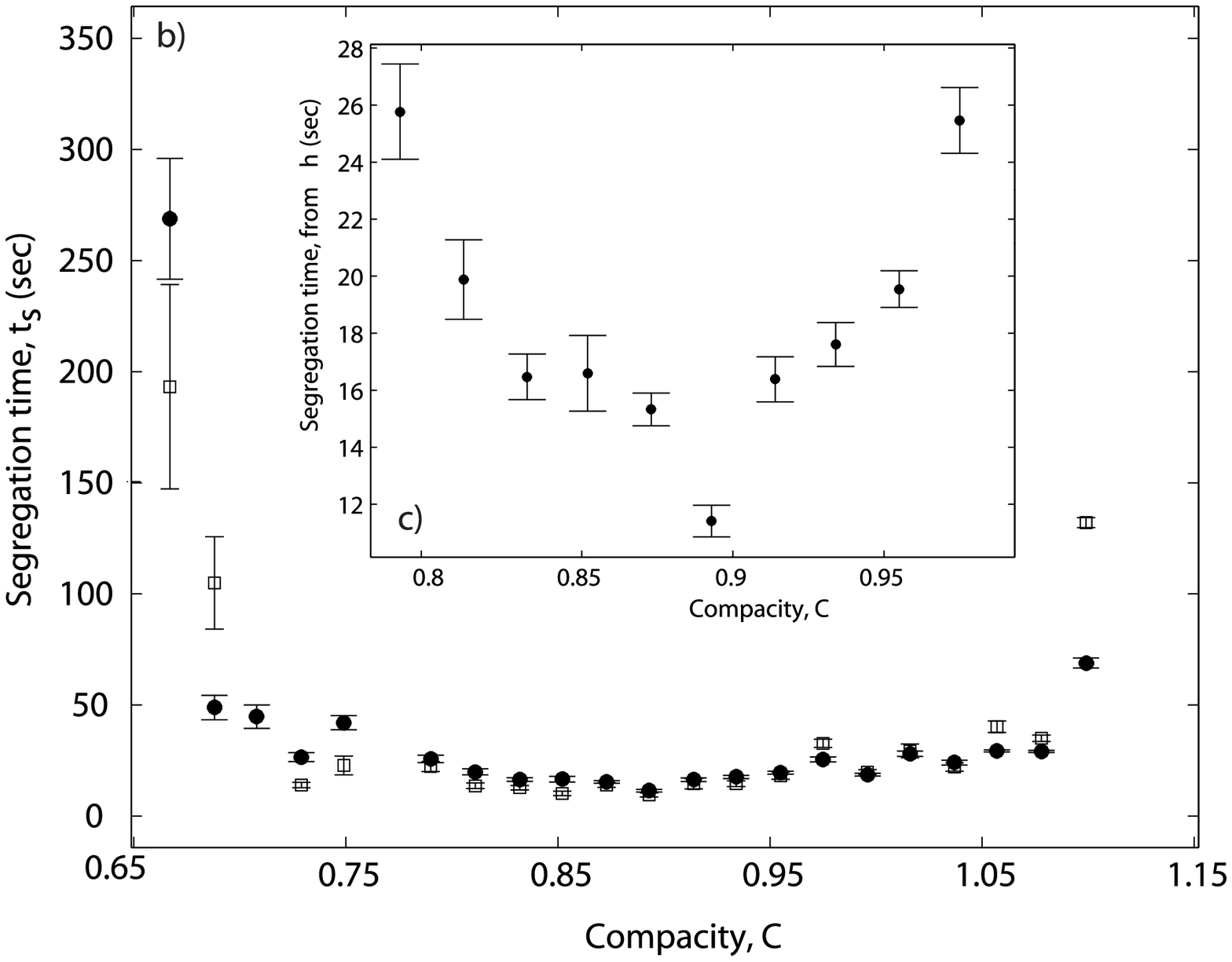}
        \end{center} \footnotesize
        \caption{ \footnotesize
        Segregation phase transition.
        (a) Compacity dependence of saturation levels from
        average domain width ($\bullet$) and average number of domains
        ($\square$). (b) Segregation time, $t_s$. from average domain width
        ($\square$) and average number of domains ($\bullet$).
        (c) Inset: Segregation time from average number of domains
        zoomed around $C=0.893$. Lines are guides to the eye.}
        \label{fig:transition}
    \end{figure}

The $C$-dependence of the segregation level is presented in Fig.
\ref{fig:transition}(a). For low compacities, up to a value,
$\mathcal{C}_c \sim 0.647$, the segregation level remains low and
approximately constant around $(3.25~\pm~0.24)mm$. This
corresponds to approximately two sphere diameters and we do not
classify these as domains; the system is deemed to be in a mixed
or unsegregated state. This is essentially a collisional regime
since there is enough free volume in the system so that particles
collide and randomly diffuse, without aggregation. We denote this
regime by a \emph{binary gas}.

As $C$ is increased, past $C_c$, clusters of the larger particles
form mobile segregation domains, of increasing $\phi$, emerge in a
nonlinear way. Above $C_c$, the clusters of larger particles were
mobile and their movement was reminiscent of oil drops on water as
they flowed, merged and split and the motion of the particles
within the clusters was highly agitated. As $C$ was incrementally
increased there was a gradual decrease of the overall motion of
segregation domains so that they become less mobile and
increasingly more stable. Also, for larger values of $C$, there
was a significant decrease of the agitation of the spheres within
the domains.

The average number of domains, shown in Fig. \ref{fig:transition}
remains approximately constant up to the same value $C_c$, after
which it rapidly decreases as $C$ is increased. This is expected
for increasingly wider domains since the number of phosphor-bronze
spheres is constant. This scenario is consistent with the
existence of a continuous phase transition for granular
segregation of the binary mixture, as the layer compacity is
incrementally increased, with a critical point for segregation,
$C_c\simeq0.647$.

In Fig. \ref{fig:transition}(b) we present the segregation times
obtained from the fits of Eq. (\ref{eqn:fits}) to $\phi(t)$ and
$\eta(t)$ as a function of $C$. Note that we only define $t_s$ for
$C>C_c$, since this is the parameter range over which segregation
occurs. As expected for a continuous phase transition, $t_s$
diverges, as $C$ is decreased from above, near $C_c$, a phenomena
usually referred to as \emph{critical slowing
down}\cite{chaikin:1995}. This adds further evidence for the
existence of a phase transition.

As the compacity is increased, past $C_c$, this slowing down of
the segregation persists up to $C=0.893$. At this point there is a
minimum in the segregation time, as shown in Fig.
\ref{fig:timeseries}(c), with $t_s=11.41s$. For greater values of
$C$ there is a qualitative change in the segregation behaviour and
the layer takes increasingly longer to segregate. At this large
values of $C$, stable crystalline stripes form which are separated
by dense regions of poppy seeds.

The above scenario suggests that the segregated phase, for
$C>C_c$, can be sub-divided into two further regimes: one in which
the speed of segregation increases with $C$ ($0.647<C<0.893$) and
another in which the opposite occurs ($C>0.893$). We denote the
first by \emph{segregation liquid} and the second by
\emph{segregation crystal}.

\section{Binary gas, segregation liquid and segregation crystal phases}
\label{sec:phases}

    \begin{figure}[h]
        \begin{center}
            \includegraphics[height=6.5cm]{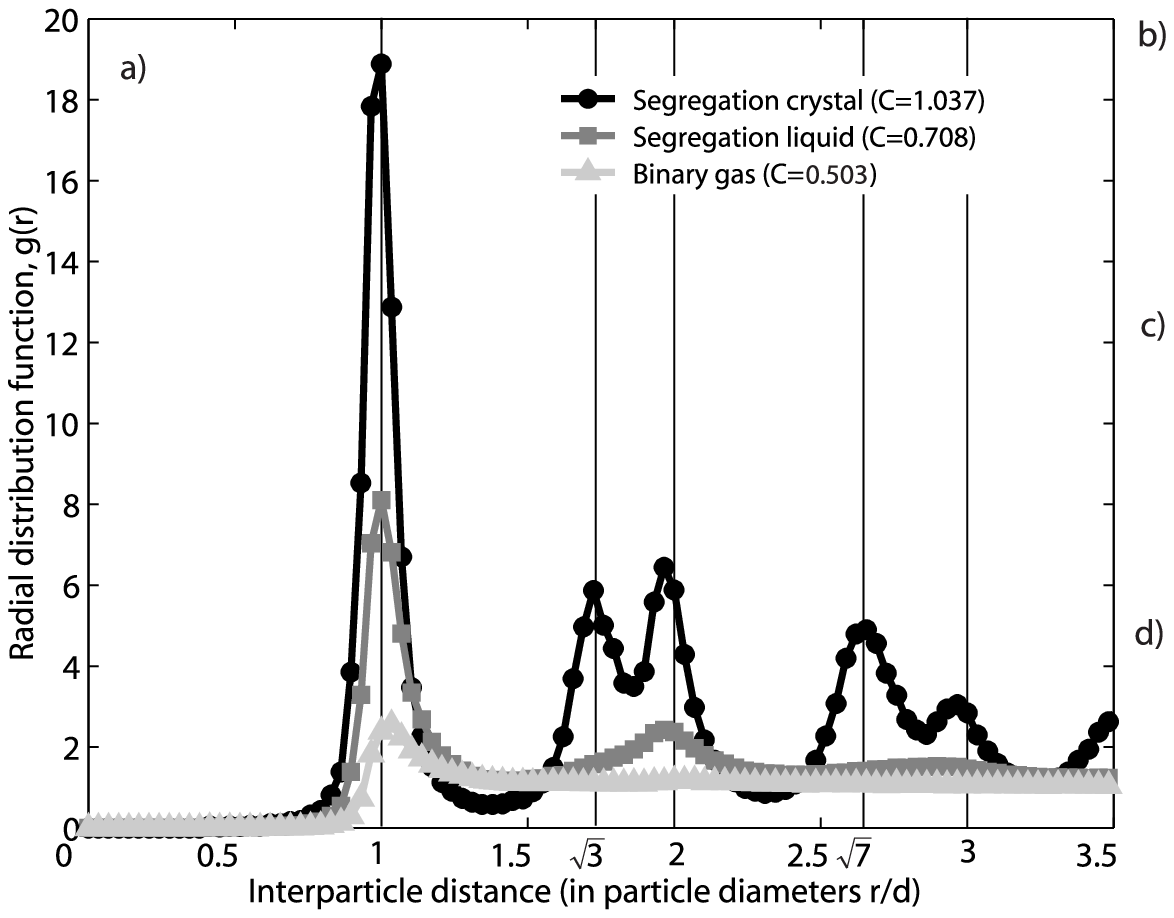}\includegraphics[height=6.5cm]{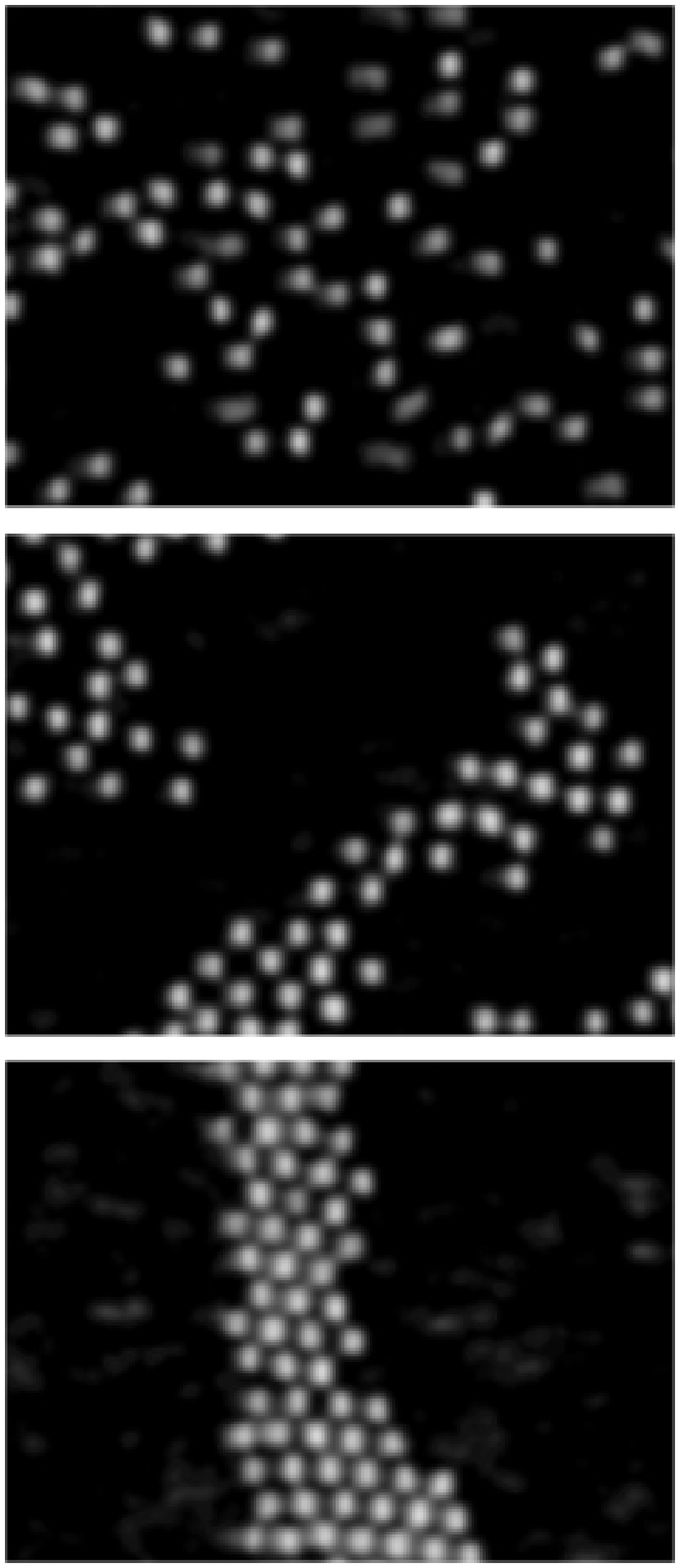}
        \end{center}
        \caption{ \footnotesize
        (a)
        Radial
distribution functions, $g(r)$, for the phosphor-bronze spheres:
($\blacktriangle$) binary gas (C=0.503), ({\tiny$\blacksquare$})
segregation liquid (C=0.708) and ($\bullet$) segregation crystal
(C=1.037). The vertical dashed lines correspond to the expected
location of the peaks for an hexagonal lattice in 2D.
        (b) Zoomed binary gas. (c) Zoomed segregation liquid. (d) Zoomed segregation
        crystal. The photographs have been image processed such that only the
        phosphor-bronze spheres are visible.}
        \label{fig:gr}
    \end{figure}

To further justify this idea of three segregation phases we turn
to the structural information given by the radial pair
distribution function for the large spheres, $g(r)$,
\begin{equation}
g(r)=A(r)\langle\sum_i\sum_{j\neq i} \delta (r-r_{ij})\rangle,
\end{equation}
where $r_{ij}$ is the separation of the $i$ and $j$th particles
and the angled brackets denote a time average. $A(r)$ is a
normalization constant such that $g(r)=1$ for a uniform
distribution of particles. The data for the calculation of $g(r)$
was obtained 3min after the driving was switched on. Time
averaging was then performed over 1min, for 1500 temporal
realisations of the layer's configuration.

In Fig. \ref{fig:gr}(a) we show three curves of $g(r)$ where each
represents the segregation phases discussed above. Typical
configurations of the spheres, for each of the phases, are
presented in the zoomed photographs in Fig. \ref{fig:gr}(b),(c)
and (d). Note that the photographs have been image processed such
that only the phosphor-bronze spheres are visible. For the binary
gas ($C<0.647$), $g(r)$ has a peak at $r/d=1$ which quickly decays
at large distances, as expected for a disordered gas. Increasing
$C$ results in a monotonic increase of the height of the first
peak, $g(d)$. At intermediate compacities ($0.647<C<0.893$),
\emph{liquid-like} behavior is observed, with $g(r)$ peaked at 1,
2 and 3 particle diameters. The positions of neighbouring spheres
are correlated and the maxima may be associated with shells of
neighbours. The oscillations are rapidly damped, showing the decay
of short-range order. This behaviour is commonly seen in hard
sphere liquids and was first observed experimentally by Bernal
\cite{bernal:1964}. In the segregation crystal phase ($C>0.893$)
two further peaks emerge near $r/d=\sqrt{3}$ and $r/d=\sqrt{7}$,
characteristic of a hexagonally packed crystal, in two dimensions.
At these high compacities, individual spheres in the segregation
domains rattle inside the cage formed by its neighbours.

\section{Oscillatory states}
\label{sec:timedependence}

\begin{figure}[t]
        \begin{center}
            \includegraphics[width=15cm]{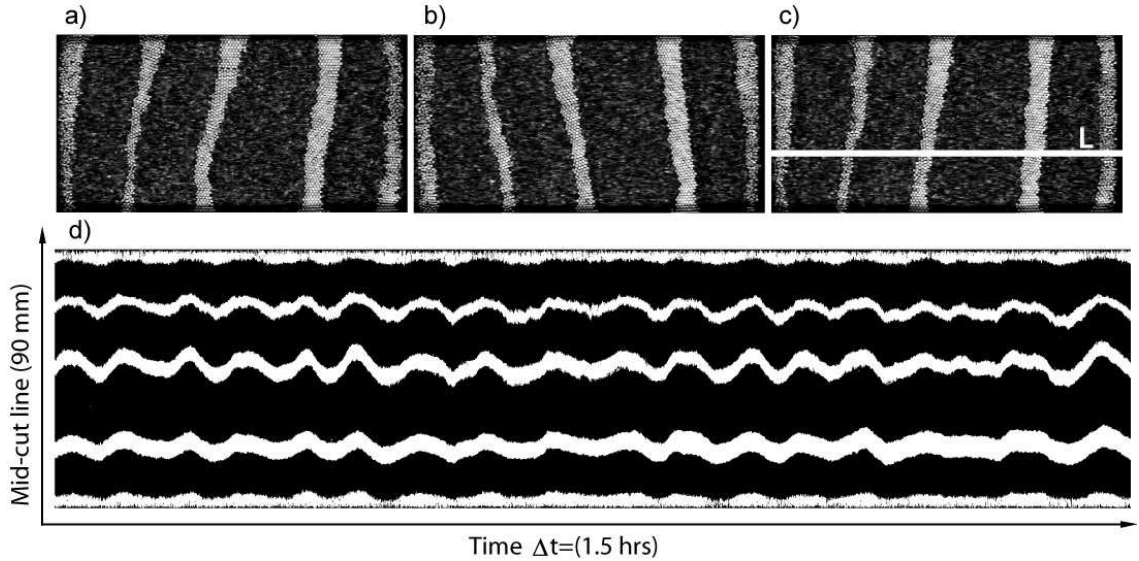}
        \end{center}
        \caption{
Photographs of the granular layer showing the oscillatory bending
of the phosphor-bronze stripes (regions in white), at different
times: (a) beginning of cycle, (b) half-cycle, (c) end of cycle.
(d) Space time diagram constructed from stacks of the cut line, L
in (c), over $1.5hrs$.
        }
        \label{fig:oscillations}
    \end{figure}

At the highest values of the compacity where robust crystalline
stripes form, we have observed intriguing well defined
oscillations of the segregation patterns. The phosphor-bronze
stripes bend, periodically, backwards and forwards in a
reproducible way. For each of the regions of poppy seeds in
between two phosphor-bronze stripes, there is a complementary
collective motion in the form of a single large vortex. The
direction of rotation of the vortex is reversed every half-cycle
of the oscillation. In Fig. \ref{fig:oscillations}(a)-(c) we show
three photographs of the granular layer which were captured 40 min
after having started from a homogenous mixture and correspond to
the beginning of the cycle, half-cycle and full cycle points of
one oscillation, respectively.

We illustrate the periodic nature of the oscillations, in Fig.
\ref{fig:oscillations}(b) with a space time diagram of the
process. This was constructed by sampling along a single line in
the $x$-dimension, and progressively stacking the lines over a
period of $1.5hrs$. The sampling line was positioned at 1/3 of the
y-length of the tray as indicated by the solid white line
superposed in Fig. \ref{fig:oscillations}(c). A well defined cycle
is clearly observed and it has a period of $T=286s$. This suggests
the existence of a strong coupling between the oscillation of the
stripes through bending and the vortex inversion in the poppy
seed's layer. It is interesting to note that the timescale
associated with these oscillations is $\sim 3500$ times slower
than the driving (at $12Hz$) excluding the possibility of a simple
resonance in the driving/granular-layer system. Hence, the
mechanism which gives rise to this oscillatory phenomena remains,
as yet, unexplained.

\section{Discussion and conclusion}
\label{sec:conclusion}

It is interesting to note that there are some analogies between
segregation in our binary granular system and aggregation in
binary colloids \cite{dinsmore:1995} and emulsions
\cite{bibette:1990}. In particular, when the size ratio between
the large colloidal particles and the radius of gyration of the
polymer is larger than $\sim0.3$, colloid-polymer mixtures display
colloidal gas, liquid and crystal phases of the large spheres
\cite{poon:2002,moussaid:2001}. Aggregation in such systems is
usually explained using an excluded volume depletion argument
introduced by Asakura and Oosawa \cite{asakura:1954} in the
context of binary hard spheres. This entropic argument is
equivalent to the mechanistic view that if two large particles are
close enough so that no other particle (or polymer) may fit
between them, they will be subjected to an asymmetric osmotic
pressure that leads to an effective attractive inter-particle
force. In colloid-polymer mixtures the strength of the interaction
can be tuned by changing the concentration of the polymer in
solution, therefore inducing gas-to-liquid and liquid-to-crystal
transitions. These resemble the granular phases we have observed
by incremental increase of the compacity. However, an important
difference is that our granular system is both driven and
dissipative and therefore far from equilibrium. In addition,
Brownian motion is an intrinsic part of the equilibrium dynamics
of colloidal systems. Hence a direct connection between the two
systems remains an open question.

In conclusion, our results provide quantitative evidence for the
existence of three phases - binary gas, segregation liquid and
segregation crystal - in granular segregation of horizontally
excited binary mixtures. At high compacities, intriguing regular
oscillatory states were observed. We show that segregation
undergoes a continuous phase transition that occurs for
compacities above a critical value and exhibits critical slowing
down.  We have presented both macroscopic and microscopic measures
which are self-consistent. Analogies of this phase behaviour may
be drawn with other depletion driven self-assembling binary
systems. This raises the possibility of importing ideas from
binary colloids in equilibrium to formulate new models for
granular segregation.
\\ \ \\

\footnotesize{The authors would like to thank A. Stephenson for
valuable discussions. P.M.R. was supported by the Portuguese
Foundation of Science and Technology. G.E. was funded by a
studentship from the EPSRC. The research of T.M. is supported by
an EPSRC Senior Fellowship. This work was completed while visiting
the Isaac Newton Institute, Cambridge.}

\bibliographystyle{unsrt}

\end{document}